\documentclass{aastex62}

\usepackage{CJK}
\usepackage[whole]{bxcjkjatype} 
\usepackage{CJKutf8}

\shorttitle{AASTeX v6.3.1 Spectropolarimetry of SN\,2021rhu}
\shortauthors{Yang et al.}
\graphicspath{{./}{figures/}}

\begin{document} 
\title{\bf Spectropolarimetry of the Thermonuclear Supernova 2021rhu: \linebreak High Calcium Polarization 79 Days After Peak Luminosity}

\correspondingauthor{Yi Yang}
\email{yiyangtamu@gmail.com}
\author[0000-0002-6535-8500]{Yi Yang
\begin{CJK}{UTF8}{gbsn}
(杨轶)
\end{CJK}}
\affiliation{Department of Astronomy, University of California, Berkeley, CA 94720-3411, USA} 
\affiliation{Bengier-Winslow-Robertson Postdoctoral Fellow}

\author[0000-0003-2560-8066]{Huirong Yan}
\affiliation{Deutsches Elektronen-Synchrotron (DESY), Platanenallee 6, 15738 Zeuthen, Germany}
\affiliation{Institut f\"ur Physik und Astronomie, Universit\"at Potsdam, Haus 28, Karl-Liebknecht-Str.\ 24/25, 14476 Potsdam, Germany}

\author[0000-0001-7092-9374]{Lifan Wang}
\affiliation{George P.\ and Cynthia Woods Mitchell Institute for Fundamental Physics \& Astronomy, Texas A\&M University, 4242 TAMU, College Station, TX 77843, USA}

\author[0000-0003-1349-6538]{J. Craig Wheeler}
\affiliation{Department of Astronomy, University of Texas, 2515 Speedway, Stop C1400, Austin, TX 78712-1205, USA}

\author[0000-0003-1637-9679]{Dietrich Baade}
\affiliation{European Organisation for Astronomical Research in the Southern Hemisphere (ESO), Karl-Schwarzschild-Str.\ 2, 85748 Garching b.\ M{\"u}nchen, Germany}

\author[0000-0002-0531-1073]{Howard Isaacson}
\affil{Department of Astronomy, University of California, 
Berkeley, CA 94720-3411, USA}

\author[0000-0001-7101-9831]{Aleksandar Cikota}
\affiliation{European Organisation for Astronomical Research in the Southern Hemisphere (ESO), Alonso de Cordova 3107, Vitacura, Casilla 19001, Santiago de Chile, Chile}

\author[0000-0003-0733-7215]{Justyn R. Maund}
\affiliation{Department of Physics and Astronomy, University of Sheffield, Hicks Building, Hounsfield Road, Sheffield S3 7RH, UK}

\author[0000-0002-4338-6586]{Peter Hoeflich}
\affiliation{Department of Physics, Florida State University, Tallahassee, Florida 32306-4350, USA}

\author[0000-0002-0537-3573]{Ferdinando Patat}
\affiliation{European Organisation for Astronomical Research in the Southern Hemisphere (ESO), Karl-Schwarzschild-Str.\ 2, 85748 Garching b.\ M{\"u}nchen, Germany}

\author[0000-0002-8965-3969]{Steven Giacalone}
\affil{Department of Astronomy, University of California Berkeley, Berkeley, CA 94720-3411, USA}

\author[0000-0002-7670-670X]{Malena Rice}
\affiliation{Department of Astronomy, Yale University, New Haven, CT 06511, USA}
\altaffiliation{NSF Graduate Research Fellow}

\author[0000-0003-0298-4667]{Dakotah B. Tyler}
\affil{Department of Physics \& Astronomy, University of California, Los Angeles, CA 90095, USA}

\author[0000-0001-5965-0997]{Divya Mishra}
\affiliation{George P.\ and Cynthia Woods Mitchell Institute for Fundamental Physics \& Astronomy, Texas A\&M University, 4242 TAMU, College Station, TX 77843, USA}
\affiliation{European Organisation for Astronomical Research in the Southern Hemisphere (ESO), Karl-Schwarzschild-Str.\ 2, 85748 Garching b.\ M{\"u}nchen, Germany}

\author[0000-0002-5221-7557]{Chris Ashall}
\affiliation{Department of Physics, Virginia Tech, Blacksburg, VA 24061, USA}

\author[0000-0001-5955-2502]{Thomas~G.~Brink}
\affiliation{Department of Astronomy, University of California, Berkeley, CA 94720-3411, USA}

\author[0000-0003-3460-0103]{Alexei~V.~Filippenko}
\affiliation{Department of Astronomy, University of California, Berkeley, CA 94720-3411, USA}

\author[0000-0002-1296-6887]{Ll\'{i}us Galbany} 
\affiliation{Institute of Space Sciences (ICE, CSIC), Campus UAB, Carrer de Can Magrans, s/n, E-08193 Barcelona, Spain}
\affiliation{Institut d’Estudis Espacials de Catalunya (IEEC), E-08034 Barcelona, Spain} 

\author[0000-0002-1092-6806]{Kishore C. Patra}
\affiliation{Department of Astronomy, University of California, Berkeley, CA 94720-3411, USA}
\affiliation{Nagaraj-Noll-Otellini Graduate Fellow}

\author[0000-0002-9301-5302]{Melissa Shahbandeh}
\affiliation{Department of Physics, Florida State University, Tallahassee, Florida 32306-4350, USA}

\author[0000-0002-4951-8762]{Sergiy~S.Vasylyev}
\affiliation{Department of Astronomy, University of California, Berkeley, CA 94720-3411, USA}
\affiliation{Steven Nelson Graduate Fellow}

\author[0000-0001-8764-7832]{Jozsef Vink{\'o}}
\affiliation{Department of Astronomy, University of Texas, 2515 Speedway, Stop C1400, Austin, TX 78712-1205, USA}
\affiliation{ Konkoly Observatory,  CSFK, Konkoly-Thege M. \'ut 15-17, Budapest, 1121, Hungary}
\affiliation{ELTE E\"otv\"os Lor\'and University, Institute of Physics, P\'azm\'any P\'eter s\'et\'any 1/A, Budapest, 1117 Hungary}
\affiliation{Department of Optics \& Quantum Electronics, University of Szeged, D\'om t\'er 9, Szeged, 6720, Hungary} 


\begin{abstract}
We report spectropolarimetric observations of the Type Ia supernova (SN) 2021rhu at four epochs: $-$7, +0, +36, and +79 days relative to its $B$-band maximum luminosity. A wavelength-dependent continuum polarization peaking at $3890 \pm 93$\,\AA\ and reaching a level of $p_{\rm max}=1.78\% \pm 0.02$\% was found. The peak of the polarization curve is bluer than is typical in the Milky Way, indicating a larger proportion of small dust grains along the sightline to the SN. After removing the interstellar polarization, we found a pronounced increase of the polarization in the \ion{Ca}{2} near-infrared triplet, from $\sim 0.3$\% at day $-$7 to $\sim 2.5$\% at day +79. No temporal evolution in high-resolution flux spectra across the \ion{Na}{1}\,D and \ion{Ca}{2}\,H\&K features was seen from days $+$39 to $+$74, indicating that the late-time increase in polarization is intrinsic to the SN as opposed to being caused by scattering of SN photons in circumstellar or interstellar matter. We suggest that an explanation for the late-time rise of the \ion{Ca}{2} near-infrared triplet polarization may be the alignment of calcium atoms in a weak magnetic field through optical excitation/pumping by anisotropic radiation from the SN. 
\end{abstract}

\keywords{polarization --- galaxies: individual (NGC 7814) --- supernovae: individual (SN\,2021rhu)}

\section{Introduction} \label{sec:intro}
The mechanism that destroys carbon/oxygen white dwarfs (CO WDs) in binary 
or multiple systems and produces Type Ia supernovae (SNe) remains 
uncertain. A key to the distinction between various possible explosion 
models is offered by measuring the explosion geometry through 
spectropolarimetry (see \citealp{Howell_etal_2001}, and the review 
by \citealp{Wang_wheeler_2008}). The continuum polarization observed in 
normal Type Ia SNe is generally low within weeks after the SN explosion; 
for instance, $p \lesssim 0.2$\% from about two weeks before
(SN\,2018gv, \citealp{Yang_etal_2020}) to six weeks after
(SN\,2001el, \citealp{Wang_etal_2003_01el, Kasen_etal_2003}; 
SN\,2006X, \citealp{Patat_etal_2009}) the peak of the optical light-curve 
maximum. This implies high overall spherical symmetry for Type Ia SNe. 
For example, when seen equator-on, an oblate ellipsoid would have an 
axis ratio of $\lesssim 1.1$ \citep{Hoeflich_1991}.  In contrast, certain 
prominent spectral lines, such as \ion{Si}{2} and \ion{Ca}{2}, usually 
show higher degrees of polarization, within weeks after the SN 
explosion \citep{Wang_wheeler_2008, Cikota_etal_2019}. This line 
polarization can be understood as the effect of clumps of chemically 
distinguished material partially blocking the underlying photosphere. 
The uneven distribution of the corresponding line opacity leads to an 
incomplete cancellation of electric ($E$) vectors over the range of the 
absorption wavelength, and hence a net line polarization. 

Ignition of an exploding WD can, in principle, be in the center, off-center, 
or throughout the volume of the WD; also, it may occur at a single knot, 
within a confined region, or at multiple locations in the WD (see overviews 
by \citealp{Alsabti_etal_2017}). Different explosion mechanisms can shape 
distinct explosion geometries and the distributions of various elements
that affect the line polarization. Here we summarize some scenarios to 
account for typical Type Ia SNe that are discussed in the literature. 

(i) Deflagration-to-detonation transition may be caused by turbulence in the 
flame front (delayed-detonation models, \citealp{Khokhlov_etal_1991}) or 
strong pulsations of the WD (pulsational delayed-detonation 
models \citealp{Khokhlov_etal_1991_pddt, Khokhlov_etal_1993,
Hoeflich_etal_1996, Plewa_etal_2004, Jordan_etal_2008, 
Bravo_etal_2009_ignition, Bravo_etal_2009_explosion, Jordan_etal_2012}). The 
abundance distribution of the burning products in the deflagration phase is 
likely to be mixed sufficiently to produce a high degree of homogeneity in 
the density distribution. Only the most prominent features such as \ion{Si}{2} 
and \ion{Ca}{2} are expected to show  polarization (of $\lesssim1$\%). The 
amplitude of line polarization may depend on the manner in which the ignition 
is initiated \citep{Seitenzahl_etal_2013, Bulla_etal_2016b}. 
(ii) Dynamical mergers between two WDs (see, e.g., \citealp{Iben_Tutukov_1984, 
Webbink_1984, Pakmor_etal_2010}) will show significant asymmetry in all 
abundances and in the density distribution, resulting in a wealth of 
significantly polarized lines ($\sim 1$\%) across the optical spectrum 
\citep{Pakmor_etal_2012, Bulla_etal_2016a}. 
(iii) Head-on collision of the WDs may exhibit bimodal chemical distributions 
including two distinct $^{56}$Ni regions \citep{Dong_etal_2015_snia} and 
significant polarization owing to strong departure from spherical symmetry.
(iv) WDs with a helium shell may explode with mass below the Chandrasekhar 
mass ($M_{\rm Ch}$) limit \citep{Taam_1980, Fink_etal_2010}. In this picture, 
a detonation in the helium layer will send a shock wave inward, triggering a 
second, off-center detonation in the inner C/O core \citep{Shen_etal_2010}. 
Two and three-dimensional hydrodynamic simulations suggest that an off-center 
detonation of the WD causes the core to be more compressed in one direction, 
thus producing an aspherical distribution of the intermediate-mass elements 
(IMEs; \citealp{Bulla_etal_2016b, Boos_etal_2021}) and significant line 
polarization. 

Multidimensional hydrodynamic computations of the polarization of a variety of
models of Type Ia SNe were conducted for the phase between $-$10 to $+$30 days 
relative to peak luminosity by \citet{Bulla_etal_2016a, Bulla_etal_2016b}. 
During this phase, the photosphere recedes into the layers of intermediate-mass
elements (IMEs) produced in the outer regions of typical explosion models. Line 
polarization of IMEs such as \ion{Si}{2} and \ion{Ca}{2} is associated with 
absorption features, and thus traces the deviation from spherical symmetry of 
the corresponding elements \citep{Hoeflich_1991, Wang_wheeler_2008}. The 
photosphere continues to recede over time. After $\sim 2$ months, most lines 
and the majority of the continuum flux come from the inner, Fe-rich ejecta. The 
polarization properties and the chemical distribution of the inner regions 
could potentially be inferred from nebular-phase spectropolarimetry; however, 
such datasets are very rare. 

Here we report spectropolarimetric observations of the Type Ia SN\,2021rhu 
between $-$7 and $+$79 days relative to the time of $B$-band maximum light. 
Two epochs of high-resolution flux spectra obtained at days $+$39 and $+$74 
in order to search for any circumstellar matter (CSM) around the SN will also 
be discussed. SN\,2021rhu was discovered  2021 July 
01 \citep[UT dates are used throughout this paper;][]{Munoz-Arancibia_etal_2021_sn2021rhu} in the alert stream of the 
Zwicky Transient Facility \citep{Bellm_etal_2019, Graham_etal_2019_ztf} in 
the edge-on spiral galaxy NGC 7814 and has been classified as a Type Ia 
SN \citep{Atapin_etal_2021_sn2021rhu}. A detailed study of the observational 
properties of SN\,2021rhu will be presented by Patra et al.\ (in prep.). \\

\section{Observations and Data Reduction} \label{sec:obs}
\subsection{VLT FORS2 Spectropolarimetry} \label{sec:fors2}
Spectropolarimetry of SN\,2021rhu was conducted with the FOcal 
Reducer and low-dispersion Spectrograph 2 
(FORS2; \citealp{Appenzeller_etal_1998}) on UT1 (Antu) of the ESO Very Large 
Telescope (VLT) in the Polarimetric Multi-Object Spectroscopy (PMOS) mode. 
With a 1$\arcsec$-wide slit, the spectral resolving power was $R\approx440$ 
[or 13\,\AA, full width at half-maximum intensity (FWHM)] at the center of 
the wavelength range from $\sim 3500$ to 9200\,\AA. An additional 
observation on day $+$0 employed the 1200R grism with 
$R \approx 2140$ \citep[corresponding to $\sim 3$\,\AA\ FWHM around the 
\ion{Si}{2} $\lambda$6355 feature; see, e.g.,][]{Anderson_etal_2018}. In order 
to maximize the blue wavelength coverage, we decided not to use an 
order-sorting filter (the standard is GG435 with a cuton wavelength of 
4350\,\AA). The contamination by second-order light, which starts beyond two 
times the atmospheric cutoff (i.e., $2 \times 3300$\,\AA), has an almost 
negligible effect on the extraction of the true polarization signal at the red 
end, unless the source is very blue (see the Appendix 
of \citealp{Patat_etal_2010}). Observations at early phases indicate that the 
photometric and spectroscopic behavior of SN\,2021rhu resembles that of normal 
Type Ia SNe \citep{Munoz-Arancibia_etal_2021_sn2021rhu, 
Atapin_etal_2021_sn2021rhu}, so the SN was not very blue. Table~\ref{Table_pol}
assembles a log of the VLT observations and the extracted polarization 
properties of SN\,2021rhu, as discussed in Section~\ref{sec:snpol}. Details of 
the FORS2/PMOS data reduction and the derivation of the Stokes parameters 
including a debias procedure of the degree of linear 
polarization introduced by \citep{Wang_etal_1997}
can be found in the FORS2 Spectropolarimetry Cookbook and Reflex 
Tutorial\footnote{\url{ftp://ftp.eso.org/pub/dfs/pipelines/instruments/fors/fors-pmos-reflex-tutorial-1.3.pdf}}, \citet{Cikota_etal_2017_fors2}, and in Appendix~A of \citet{Yang_etal_2020}, following the procedures 
of \citet{Patat_etal_2006_polerr} and \citet{Maund_etal_2007_05bf}.

Below $\sim 4000$\,\AA, the sensitivity of the VLT FORS2 instrument decreases 
rapidly. Therefore, flux calibration and estimation of the polarization error 
at the very blue end of the optical spectrum may suffer from systematic 
uncertainties, and the polarization features become hard to characterize. 

\subsection{Keck HIRES spectroscopy} \label{sec:hires}
We obtained two sets of spectra of SN\,2021rhu with the High-Resolution Echelle 
Spectrometer (HIRES; \citealp{Vogt_etal_1994}) instrument on the Keck-I 10\,m 
telescope on 2021-08-23 (day $+$39) and 2021-09-27 (day $+$74). We used the C2 
decker setup (14\arcsec$\times$0\farcs{861}, $R = 45,000$) and integrated for 
$3 \times 900$\,s and $2 \times 1800$\,s, respectively, at the two epochs. The 
spectra were reduced following a standard routine of the California Planet 
Search project \citep{Howard_etal_2010}. We corrected the velocity zero point 
of the spectral orders containing the \ion{Na}{1}\,D and \ion{Ca}{2}\,H\&K 
lines to the rest frame using the recession velocity of 
$v_{\rm res} = 1049$\,km\,s$^{-1}$ measured for the spiral host galaxy 
NGC 7814 \citep{van_Driel_etal_2016} and the barycentric velocity determined 
following \citet{Wright_etal_2014} for the UTC at each HIRES observation. \\

\section{Results} \label{sec:results}

\begin{figure}[h]
\epsscale{0.75} 
\plotone{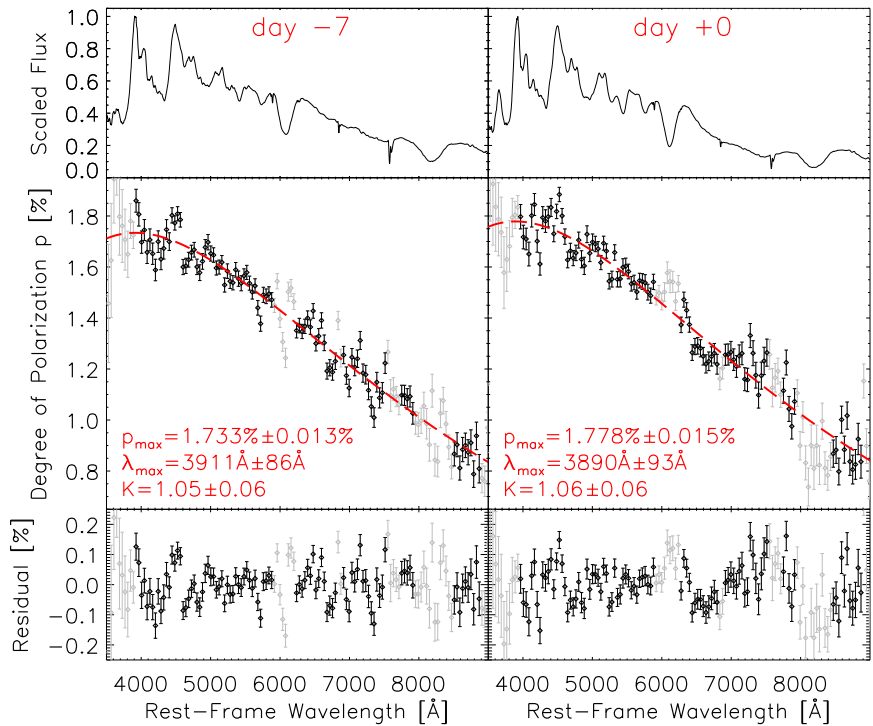}
\caption{
Estimation of the ISP toward SN\,2021rhu based on the observations on days $-$7 
(left column) and $+$0 (right column). The top row gives the flux spectra 
normalized to the maximum value within the range. The middle row presents the 
degree of linear polarization for the given epochs. The red long-dashed curve 
indicates the best fit using a Serkowski law. The fitted parameters are labeled 
in each panel. The data have been rebinned to 40\,\AA. Data points plotted in 
gray were excluded from the fitting since they cover broad and polarized features 
or major telluric lines. The bottom row shows the residuals from the fit. 
\label{Fig_isp} 
}
\end{figure}

\begin{figure}[!htb]
   \begin{minipage}[t]{0.48\textwidth}
     \centering
     \includegraphics[width=1.03\linewidth]{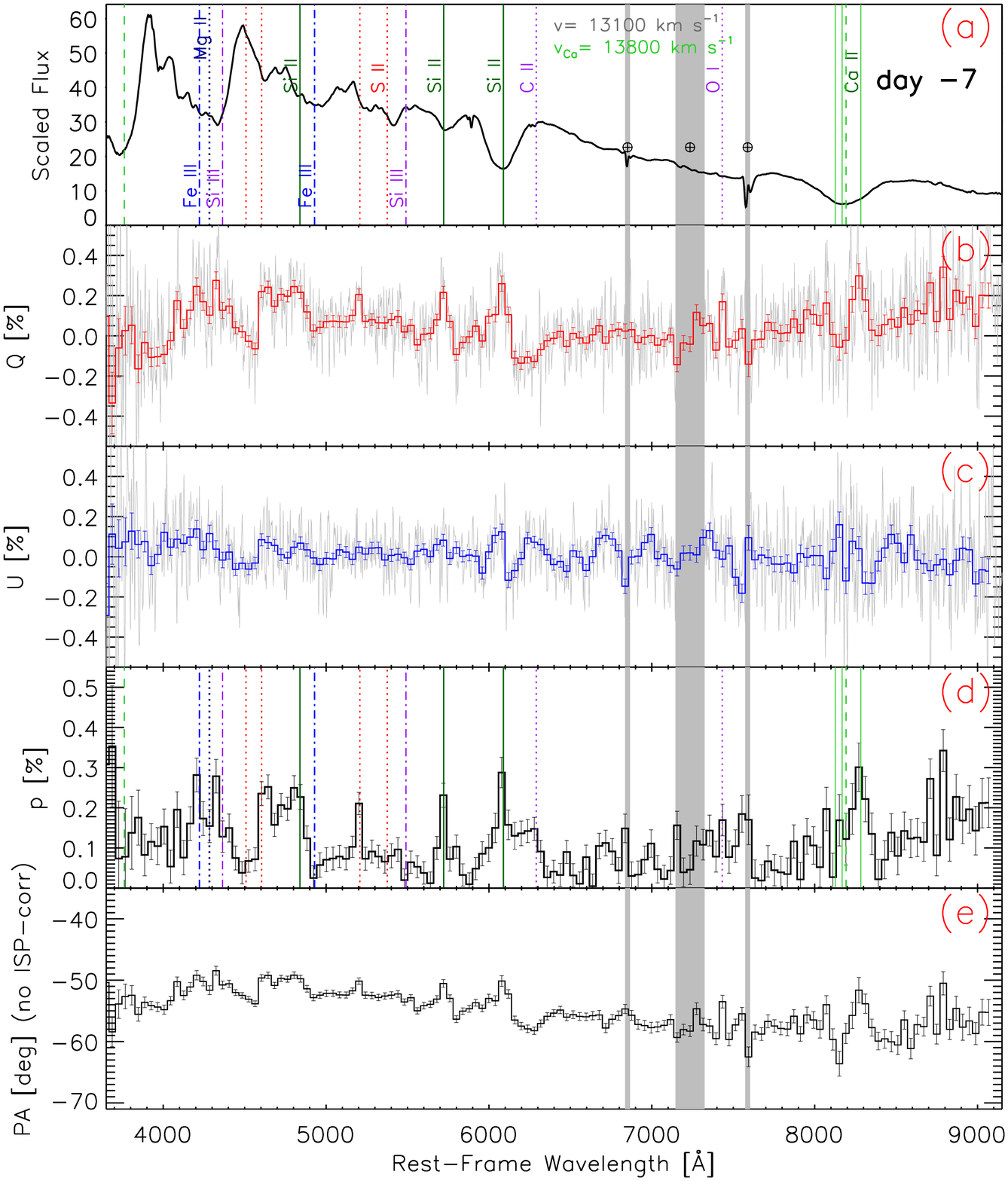}
     \caption{Spectropolarimetry of SN\,2021rhu observed on day $-$7. The five 
     panels (from top to bottom) present (a) the scaled-flux spectrum; (b,c) 
     the normalized Stokes parameters $Q$ and $U$, respectively; (d) the 
     polarization spectrum ($p$); and (e) the polarization position angle (PA). 
     The diagrams in panels (b)--(d) represent the polarization after ISP 
     correction. PAs in panel (e) are presented before subtracting the ISP. The 
     data have been rebinned to 40\,\AA, while the Stokes $Q$ and $U$ parameters 
     approximately sampled with the detectorpixel size are indicated by gray 
     histograms in panels (b)--(c). 
     }\label{Fig_iqu_ep1}
   \end{minipage}\hfill
   \begin{minipage}[t]{0.48\textwidth}
     \centering
     \includegraphics[width=1.03\linewidth]{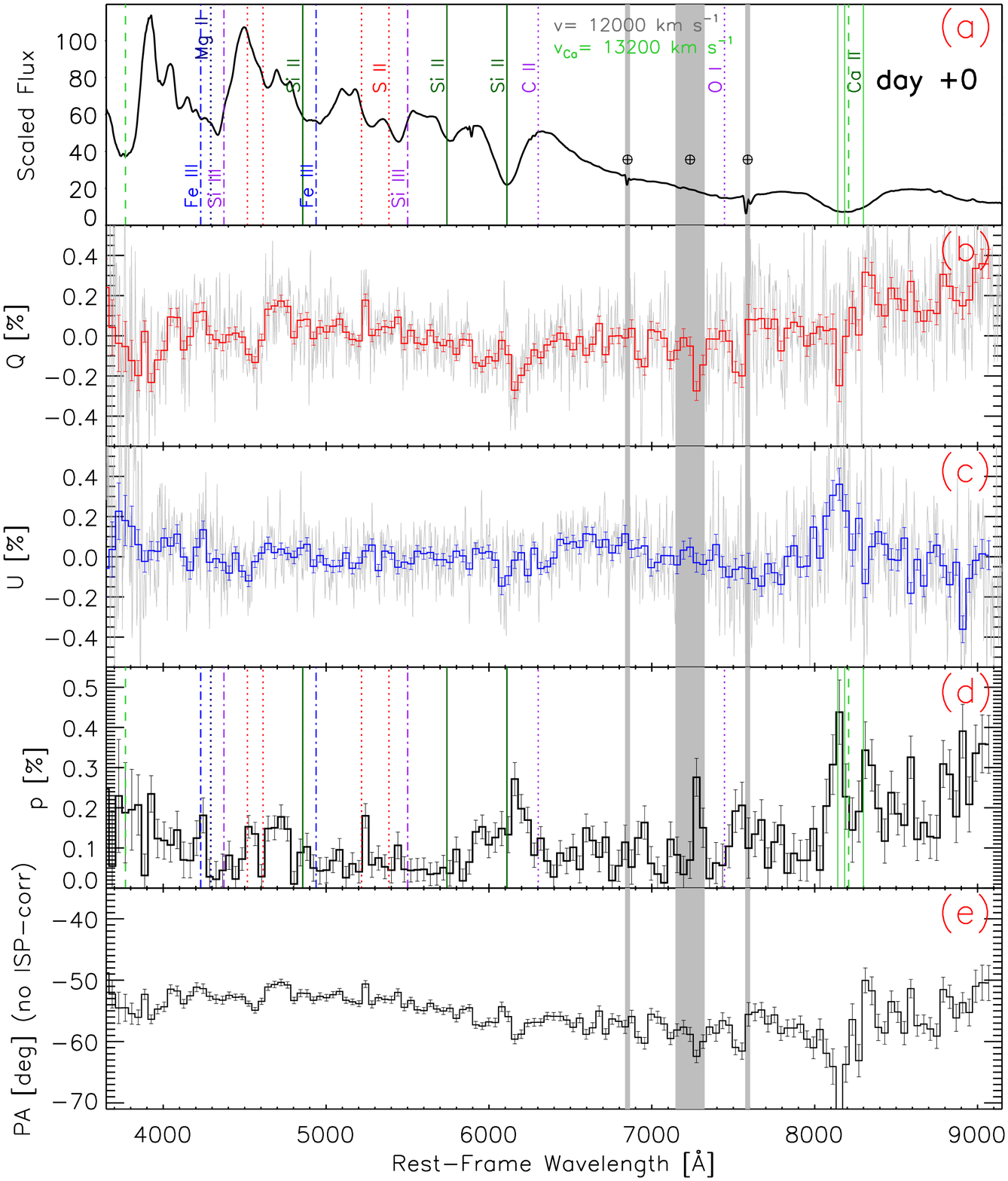}
     \caption{Same as Fig.~\ref{Fig_iqu_ep1}, but for 
     day $+$0.}\label{Fig_iqu_ep2}
   \end{minipage}
\end{figure}

\begin{figure}[!htb]
   \begin{minipage}[t]{0.48\textwidth}
     \centering
     \includegraphics[width=1.03\linewidth]{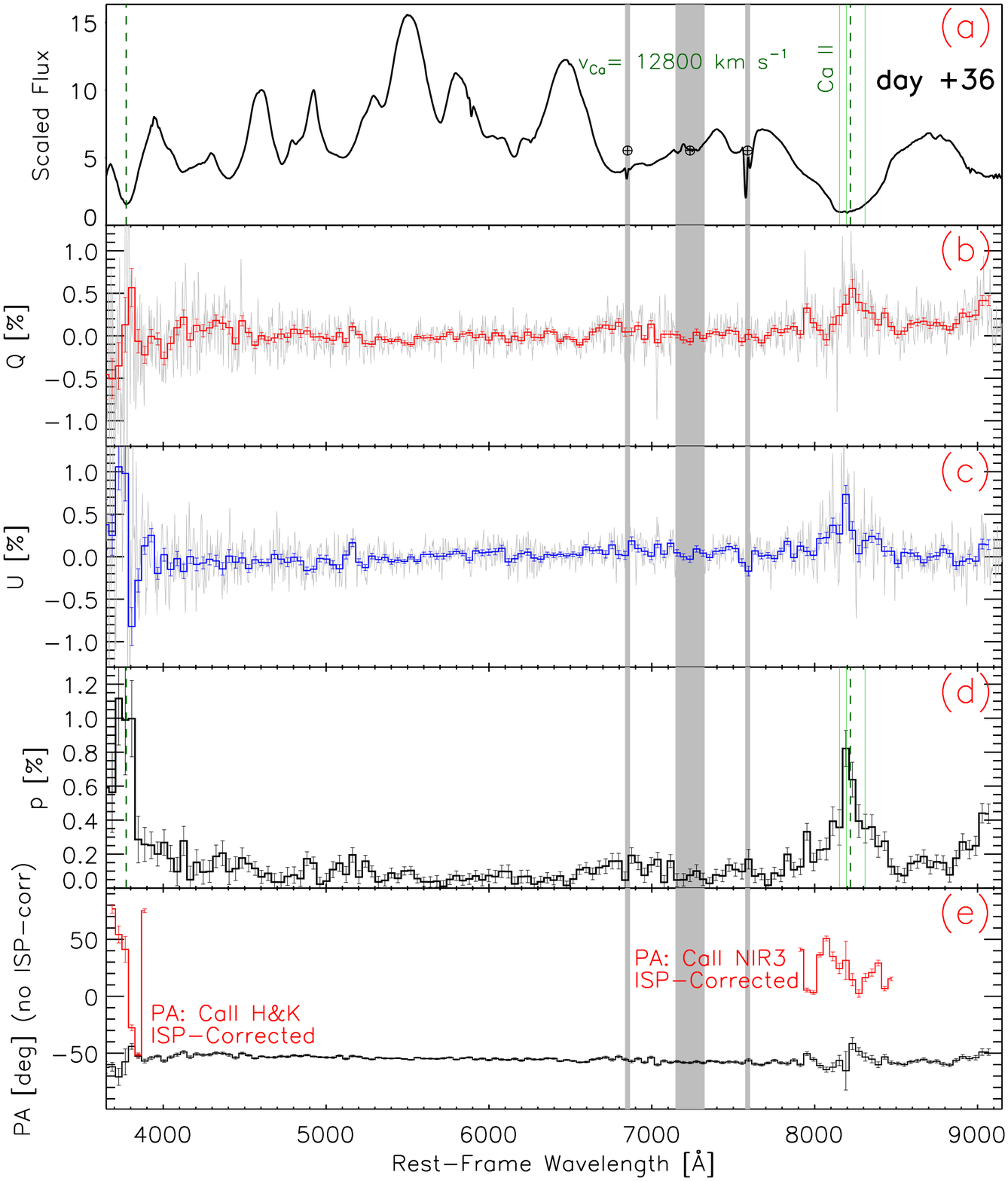}
     \caption{Same as Fig.~\ref{Fig_iqu_ep1}, but for day $+$36. The red 
     histograms in panel (e) display the ISP-corrected PA across 
     \ion{Ca}{2}\,H\&K and \ion{Ca}{2}\,NIR3 after correction for the 
     ISP.}\label{Fig_iqu_ep4}
   \end{minipage}\hfill
   \begin{minipage}[t]{0.48\textwidth}
     \centering
     \includegraphics[width=1.03\linewidth]{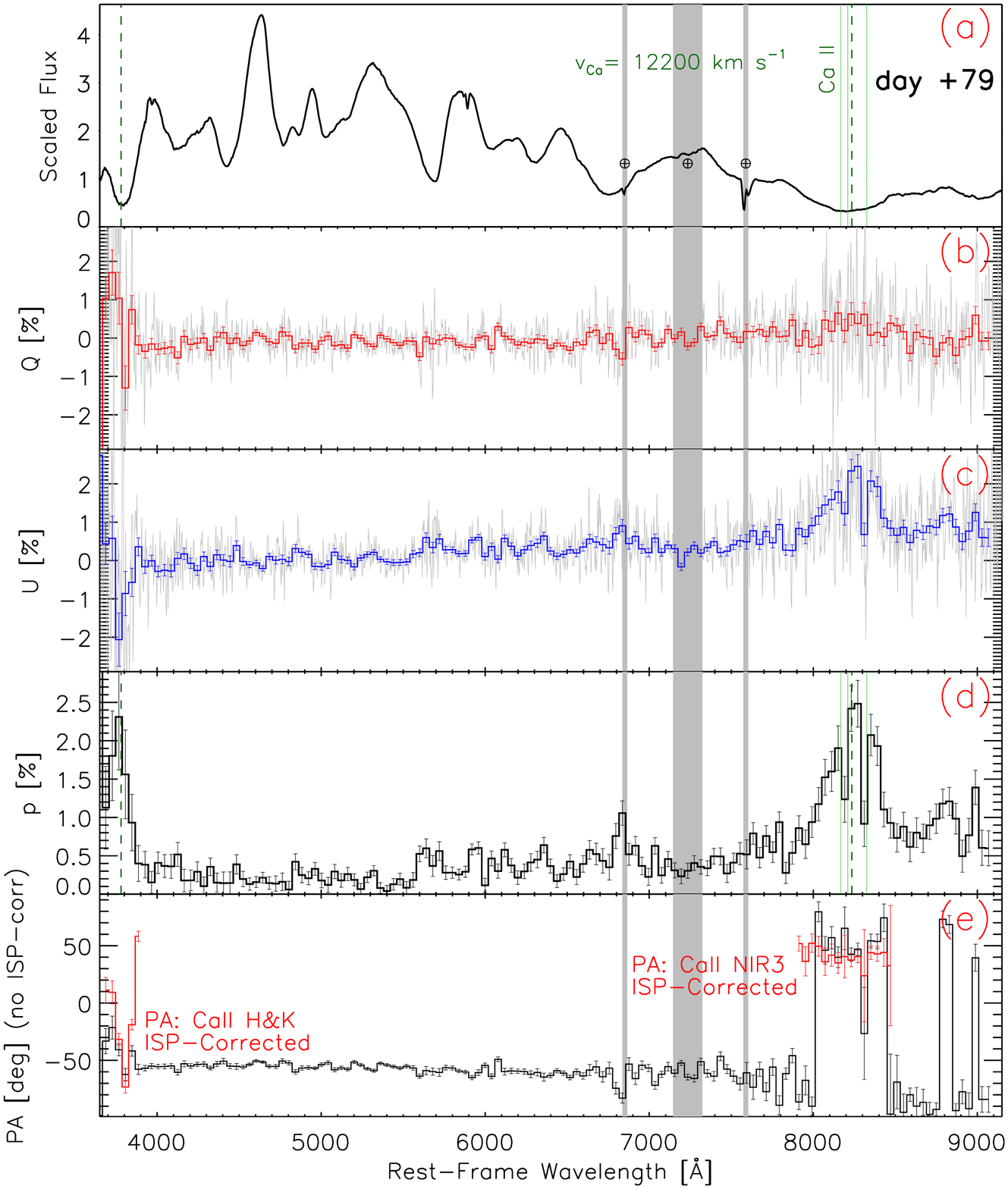}
     \caption{Same as Fig.~\ref{Fig_iqu_ep1}, but for 
     day $+$79. The red histograms in panel (e) display the ISP-corrected PA 
     across \ion{Ca}{2}\,H\&K and \ion{Ca}{2}\,NIR3 after correction for the 
     ISP.}\label{Fig_iqu_ep5}
   \end{minipage}
\end{figure}

\subsection{Interstellar Polarization} \label{sec:isp}
At all epochs, the polarization measured toward SN\,2021rhu exhibits a clear 
wavelength dependence (see the middle panels of Fig.~\ref{Fig_isp}). In the 
optical/near-infrared domain, the wavelength ($\lambda$) dependence of the 
interstellar polarization (ISP) can be approximated by the Serkowski 
law \citep{Serkowski_etal_1975},
\begin{equation}
p(\lambda)/p(\lambda_{\rm max}) = {\rm exp}[-K\ \rm {ln}^2 (\lambda_{\rm max} / \lambda)] , 
\label{Eqn_Ser}
\end{equation}
where $\lambda_{\rm max}$ and $p(\lambda_{\rm max})$ denote the wavelength 
and the level of the maximum polarization, respectively. The parameter $K$ 
describes the width of the interstellar polarization peak. For the purpose 
of estimating the ISP, we exploit the fact that the intrinsic continuum 
polarization of Type Ia SNe is generally negligible (i.e., 
$\lesssim 0.2$--0.3\%; \citealp{Wang_wheeler_2008}). Hence, we fitted 
Serkowski's law to the polarization spectra of SN\,2021rhu at days $-$7 and 
$+$0, and present the results in Figure~\ref{Fig_isp}. Data points near the 
blue end of the spectra or belonging to the prominent and polarized 
\ion{Si}{2}\,$\lambda$6355 and \ion{Ca}{2} near-infrared triplet (NIR3) 
features were excluded in the fitting process. We take the \ion{Ca}{2} NIR3 
feature to have a rest wavelength of $\lambda_{\rm0, Ca}=8570$\,\AA, averaged 
over the wavelengths of the triplets (8500.36\,\AA, 8544.44\,\AA, and 
8664.52\,\AA). The fitted ISP curve for day $+$0 ($K=1.06\pm0.06$, 
$\lambda_{\rm max} = 3890\pm93$\,\AA, and 
$p(\lambda_{\rm max})=1.778\pm0.015$\%) has been adopted for the ISP 
corrections throughout the paper. The parameters are consistent with the 
values determined based on observations at day $-$7 (see the left panels of 
Fig.~\ref{Fig_isp}), thus confirming the assumption of constant low 
continuum polarization of SN\,2021rhu around its peak luminosity. The actual 
fitting and correction process has been carried out separately for the Stokes 
parameters $Q$ and $U$ to obtain their values intrinsic to the SN at all 
observed phases. We also fitted the ISP by adding a secondary component which 
characterizes any contribution from the Galactic dust. A Galactic reddening 
component can be estimated as $E(B-V)^{\rm MW} = 0.04$\,mag based 
on \citet{Schlafly_etal_2011} and \citet{Cardelli_etal_1989}. This provides 
an upper limit on the polarization by the Milky Way dust induced by dichroic 
extinction: 
$p_{\rm ISP} \textless 9\% \times E(B-V)$ \citep{Serkowski_etal_1975}, and 
$p_{\rm max}^{\rm MW}\leq0.36$\%. Adopting a fixed 
$K^{\rm MW} = 1.15$ \citep{Serkowski_etal_1975}, we found that 
$p_{\rm max}^{\rm MW}$ is  consistent with zero, suggesting that the observed 
ISP toward SN\,2021rhu is mainly contributed by its host.

\subsection{Intrinsic Polarization of SN\,2021rhu} \label{sec:snpol}
In Figures~\ref{Fig_iqu_ep1}--\ref{Fig_iqu_ep5} we present the ISP-corrected 
spectra of Stokes parameters, degree of polarization, and polarization 
position angle obtained at four epochs from days $-$7 to $+$79. The figures 
cover a wavelength range 3600--9150\,\AA\ and the data have been rebinned 
to 40\,\AA\ in order to increase the signal-to-noise ratio but also make sure 
that major broad spectral features are sampled by at least $\sim 5$--10 
resolution elements. The polarization position angle in panel (e) is presented 
without subtracting the ISP since ${\rm PA} = (1/2){\rm tan}^{-1}(U/Q)$ will 
display random values when $Q$ is low after a baseline ISP has been removed. 
The size of the error bars in the histograms that provide the 
polarization measurements represent the 1--$\sigma$ uncertainty.

\begin{figure}[h]
\epsscale{0.7} 
\plotone{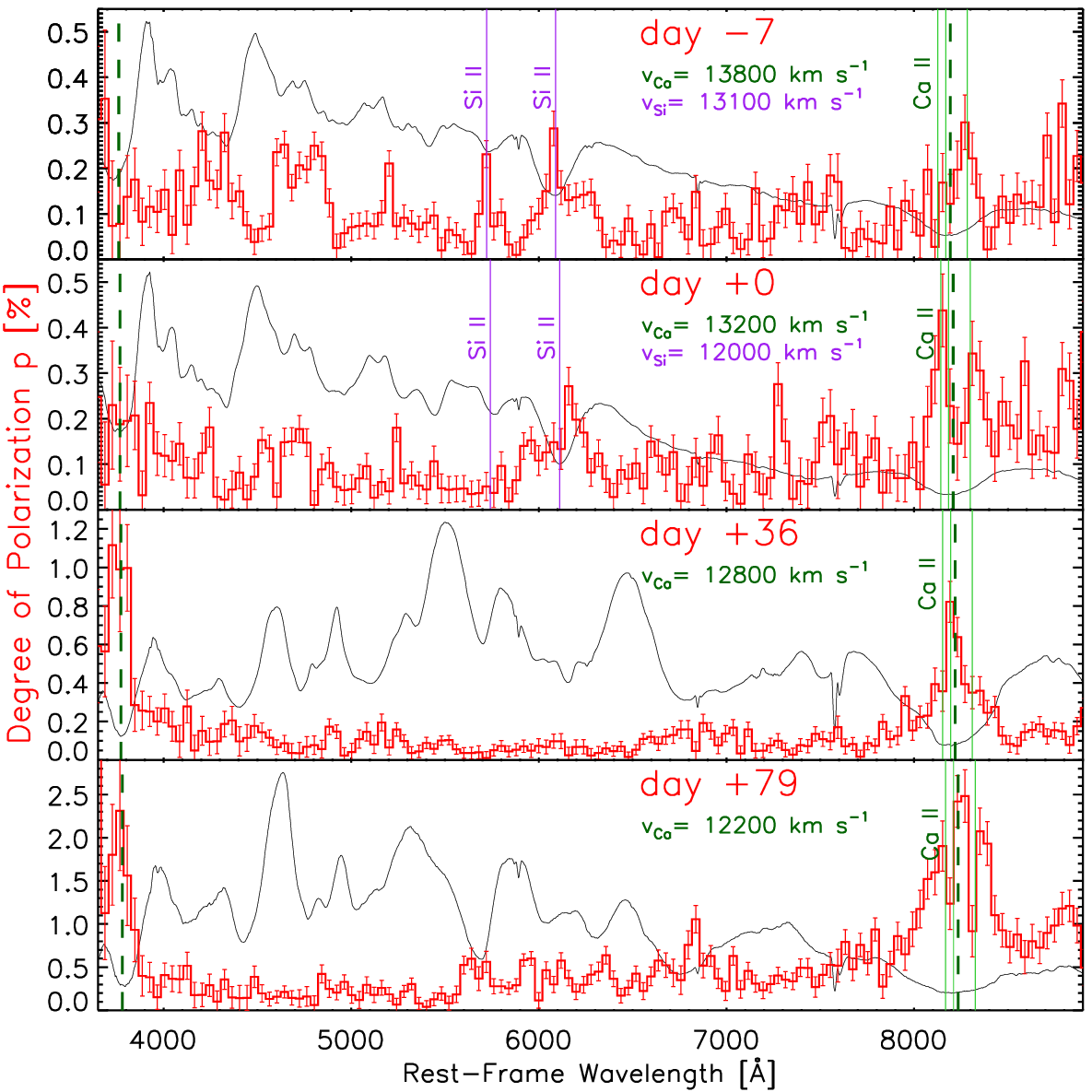}
\caption{Evolution of the intrinsic polarization of SN\,2021rhu. For each 
epoch, the degree of polarization is presented (red histogram) with 40\,\AA\ 
binning, together with the full-resolution and arbitrarily scaled flux 
spectrum (black line). \ion{Ca}{2}\,NIR3 features at all epochs are labeled 
with their velocities ($v_{\rm Ca}$) calculated with 
$\lambda_{\rm0, Ca}=8570$\,\AA\ (Sec.\,\ref{sec:isp}). Vertical light-green 
lines indicate the locations of the three components of \ion{Ca}{2}\,NIR3. 
The two vertical green-dashed lines trace the evolution of the 
\ion{Ca}{2}\,H\&K and the  \ion{Ca}{2}\,NIR3 velocities. Vertical purple 
lines identify the \ion{Si}{2}\,$\lambda$5976 and \ion{Si}{2}\,$\lambda$6355 
features at days $-7$ and $+$0. The Ca and Si velocities were determined from 
the absorption minima corresponding to $\lambda_{\rm0, Ca}$ and 
\ion{Si}{2}\,$\lambda$6355, respectively. The degree of polarization across 
both \ion{Ca}{2}\,NIR3 and \ion{Ca}{2}\,H\&K shows a substantial increase at 
later epochs. 
\label{Fig_pol}
}
\end{figure}

\begin{figure}[h]
\epsscale{0.65} 
\plotone{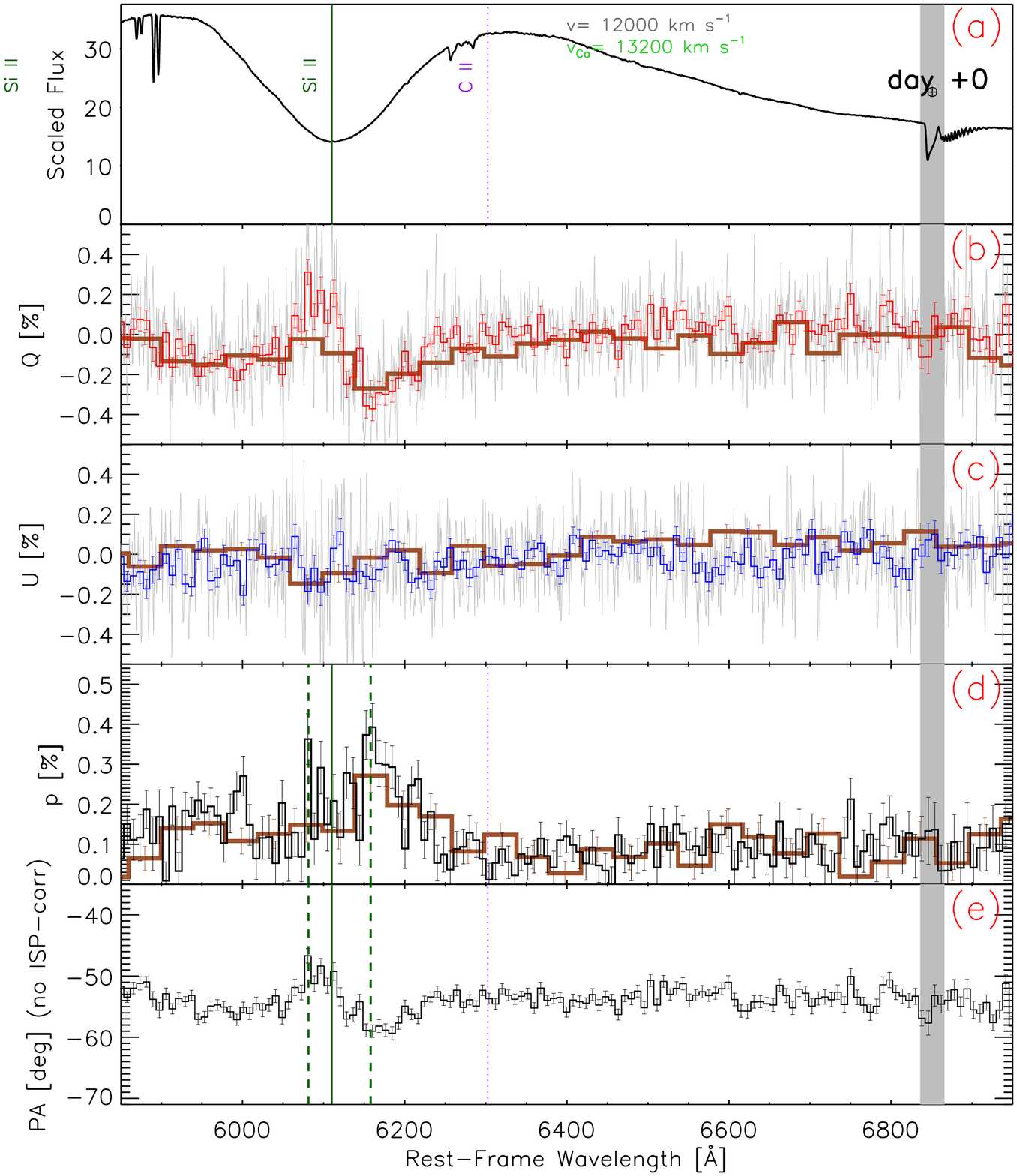}
\caption{Similar to Fig.~\ref{Fig_iqu_ep1} but for day $+$0 at higher spectral 
resolution. The data have been rebinned to 8\,\AA. For comparison, 
low-resolution data obtained at the same epoch rebinned to 40\,\AA\ are shown 
by the brown thick histograms in panels (b)--(d). The vertical solid green 
lines in panels (a) and (d) represent the minimum of the total-flux absorption 
line at 12,000\,km\,s$^{-1}$. The left and right vertical dashed lines in panel 
(d) mark the two polarization peaks at 13,500 and 9,600\,km\,s$^{-1}$, 
respectively. The gray-shaded area identifies a region with a major telluric 
feature. 
\label{Fig_iqu_ep3}
}
\end{figure} 

The temporal evolution of the degree of polarization of SN\,2021rhu after the 
ISP correction is shown in Figure~\ref{Fig_pol}. We also estimate the 
continuum polarization based on the Stokes parameters over the wavelength 
range 5000--6800\,\AA\ with the highly polarized \ion{Si}{2} lines excluded. 
The error-weighted mean Stokes parameters $(q^{\rm Cont}$, $u^{\rm Cont})$ 
across this region from days $-7$ to $+$79 are presented in 
Table~\ref{Table_pol}. The error has been estimated by adding the statistical 
uncertainties and the standard deviation calculated from the 40\,\AA\ binned 
spectra within the continuum wavelength range in quadrature. 
The $q^{\rm Cont}$ and $u^{\rm Cont}$ are consistent with zero within the 
selected wavelength range, and the level of 
polarization intrinsic to SN\,2021rhu across the optical continuum remains 
low from days $-$7 to $+$79. The most noticeable behavior in the 
spectropolarimetric evolution of SN\,2021rhu is the strong increase in the 
peak polarization of the \ion{Ca}{2} lines from days $+$36 to $+$79, namely 
in \ion{Ca}{2}\,H\&K from $1.1 \pm 0.3$\% to $2.3 \pm 0.7$\%, and in 
\ion{Ca}{2}\,NIR3 from $0.8 \pm 0.1$\% to $2.5 \pm 0.3$\% (see 
Table~\ref{Table_pol} and Fig.~\ref{Fig_pol}), as measured in the data with 
40\,\AA\ bin size. At all epochs, the peak polarization across 
\ion{Ca}{2}\,NIR3 is significantly higher than in the continuum. 

A polarization signal is seen in \ion{Si}{2}\,$\lambda$6355 on days $-$7 and 
$+$0. Thereafter, it vanished in accordance with the disappearance of the 
\ion{Si}{2} feature from the total-flux spectrum a few weeks past maximum 
light. This early \ion{Si}{2} polarization can be understood as an 
inhomogeneous obscuration of the photosphere. With the recession of the 
photosphere into theinterior ejecta below the Si-rich layer, the optical depth 
of the Si becomes small, and polarization by blocking portions of the 
photosphere cannot occur. 

The polarization profiles of the \ion{Ca}{2}\,NIR3 and 
\ion{Si}{2}\,$\lambda$6355 lines exhibit rich structures that also evolve with 
time. For the former complex, we infer the expansion velocity from the 
absorption minimum of the \ion{Ca}{2}\,NIR3 feature in the total-flux spectrum, 
namely $v_{\rm Ca}=$ 13,800, 13,200, 12,800, and 12,200\,km\,s$^{-1}$, at days 
$-7$, $+$0, $+$36, and $+$79, respectively, as shown in 
Figures~\ref{Fig_iqu_ep1}--\ref{Fig_pol}. We also mark the corresponding 
expansion velocities of the three transitions with green vertical solid lines. 
The \ion{Ca}{2}\,H\&K feature has almost the same velocity as measured from 
\ion{Ca}{2}\,NIR3, and we also characterize it with $v_{\rm Ca}$. The velocity 
of \ion{Si}{2}\,$\lambda$6355 is also labeled in 
Figures~\ref{Fig_iqu_ep1}-\ref{Fig_iqu_ep2} and \ref{Fig_pol}--\ref{Fig_iqu_ep3}, 
namely $v=$ 13,100 and 12,000\,km\,s$^{-1}$ at days $-7$ and $+$0, respectively. 
Identifications of major spectral lines are provided for days 
$-$7 and $+$0, before the SN has entered the nebular phase. All spectral features 
except for the \ion{Ca}{2} lines marked in Figures~\ref{Fig_iqu_ep1} and 
\ref{Fig_iqu_ep2} are labeled for the photospheric velocity $v$. Different colors 
were used to provide a better separation of the lines. Major telluric features 
are marked with gray-shaded areas.

At day $+$0, two peaks appear in the polarization profile of \ion{Ca}{2}\,NIR3, 
at $\sim 15,400$ and $\sim 8700$\,km\,s$^{-1}$ with respect to the rest frame 
of SN\,2021rhu. They bracket the \ion{Ca}{2}\,NIR3 and \ion{Ca}{2}\,H\&K 
velocities of $\sim 13,200$\,km\,s$^{-1}$ measured at the same epoch. By day 
$+$79, three peaks near 15,700, 11,500, and 7,500\,km\,s$^{-1}$ had developed 
in the \ion{Ca}{2}\,NIR3 polarization profile. At this epoch, $v_{\rm Ca}$ has 
decelerated less and appears at 12,200\,km\,s$^{-1}$. The uncertainty of the 
stated velocities is dominated by the width of the smallest resolution element, 
namely half of 8\,\AA\ for \ion{Si}{2}$\,\lambda$6355 in the higher-resolution 
observation and half of 40\,\AA\ for \ion{Ca}{2}\,NIR3, corresponding to 
$\sim 200$ and $\sim 700$\,km\,s$^{-1}$, respectively. The uncertainties 
represent the maximum possible error owing to rounding of wavelengths in 
spectral bins. Inspection of the raw data before spectral rebinning confirms 
the reality of the structures as well as of their evolution. From the available 
data, it is not possible to conclude whether the three polarization peaks 
correspond to the three triplet components or have a different origin. 

At days $-$7 and $+$0, the expansion velocity of the SN ejecta inferred from 
the absorption minimum in the \ion{Si}{2}\,$\lambda$6355 flux profile was 
13,100 and 12,000\,km\,s$^{-1}$ (marked by the vertical solid and dotted-dashed 
lines in Figures~\ref{Fig_iqu_ep1}--\ref{Fig_iqu_ep2} and ~\ref{Fig_pol}), 
respectively. The higher-resolution polarization profile of this line from day 
$+$0 (this observation does not cover the \ion{Ca}{2}\,NIR3 region) exhibits 
two peaks. At about 13,500 and 9,600\,km\,s$^{-1}$, respectively (highlighted 
by the dashed dark-green lines in Figs.~\ref{Fig_iqu_ep3}d 
and \ref{Fig_iqu_ep3}e), their velocities are different from that of the 
absorption minimum. A peak and a trough in $Q$ at the position of 
\ion{Si}{2}\,$\lambda$6355 (Fig.~\ref{Fig_iqu_ep3}b) lead to different 
position angles (Fig.~\ref{Fig_iqu_ep3}e). Without ISP subtraction, the 
polarization position angles of the blue ($-50.0\pm2.5$ deg) and the red 
($-57.5\pm2.0$ deg) components bracket that of the continuum 
($-53.5 \pm 1.3$ deg). The position angles of the blue and the red components 
were estimated by taking the error-weighted mean value in a velocity range of 
$\pm 800$\,km\,s$^{-1}$ relative to the respective peak. The continuum PA was 
computed between 6300\,\AA\ and 6700\,\AA. In the low-resolution observations 
on the same date with grism 300V (overplotted in brown in 
Figs.~\ref{Fig_iqu_ep3}(b)--(d)), only traces of the complex polarization 
structure are visible as some asymmetry. The differences between 
\ion{Si}{2}\,$\lambda$6355 and \ion{Ca}{2}\,NIR3 in the structure and evolution 
of their multiple polarization components will be analyzed by Patra et al.\ 
(in prep.).

\subsection{\ion{Na}{1}\,D and \ion{Ca}{2}\,H\&K Flux Profiles}
Since any CSM will be exposed to intense ultraviolet (UV) radiation from the 
SN explosion, embedded elements such as Na and Ca are likely to be ionized. 
Unlike most of the broad features that arise from the expanding ejecta which 
retain high kinetic energy, lines from these elements would be narrow enough 
so that any subcomponents of each feature could be seen separately with higher 
spectral resolution. Additionally, the \ion{Na}{1}\,D line provides a good 
tracer of gas and dust, and its strength is correlated with the dust reddening 
along the line of sight \citep{Munari_etal_1997,Poznanski_etal_2012,
Phillips_etal_2013}. Temporal variability of such narrow absorption lines 
would indicate the evolving conditions of the CSM ionization induced by the 
variable SN radiation field. Therefore, this observational signature has been 
used to search for CSM around Type Ia SNe \citep{Patat_etal_2007,
Simon_etal_2009, Sternberg_etal_2011, WangX_etal_2019}. 

\begin{figure}[h]
\epsscale{0.75} 
\plotone{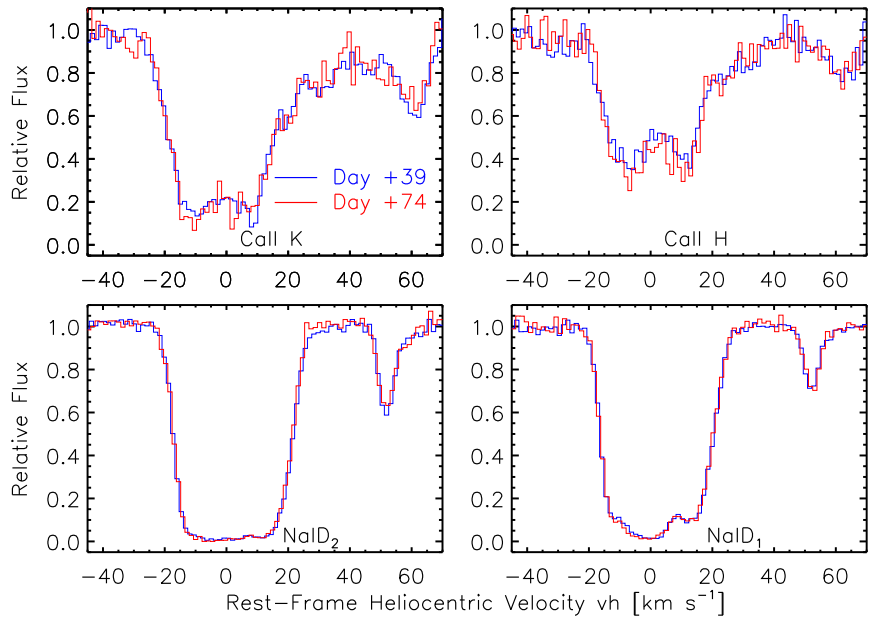}
\caption{Keck/HIRES spectra of SN\,2021rhu in velocity space relative to 
\ion{Ca}{2}\,K (upper left), \ion{Ca}{2}\,H (upper right), \ion{Na}{1}\,D$_{2}$ (lower left), and \ion{Na}{1}\,D$_{1}$ (lower right) features. The observed 
spectra have been divided by the respective ambient pseudocontinuum. No 
apparent temporal evolution between the two epochs has been identified. 
\label{Fig_hires}
}
\end{figure}

In order to investigate any temporal evolution of circumstellar Na and Ca line 
profiles of SN\,2021rhu, we approximate the pseudocontinuum by a low-order 
polynomial fitted to the spectrum between $\sim \pm 20$\,\AA\ of the central 
wavelength with the absorption line excluded. The flux spectrum spanning each 
line profile was then divided by the pseudocontinuum spectrum. The line 
profiles are shown in Figure~\ref{Fig_hires}. For SN\,2021rhu, we identify no 
temporal evolution in the \ion{Na}{1}\,D doublet (5895.92, 5889.95\,\AA) and 
the \ion{Ca}{2}\,H\&K doublet (3968.47, 3933.66\,\AA). At both epochs, two 
complexes of absorption features with centroids at $\sim -10$ and 
$\sim +10$\,km\,s$^{-1}$ plus a shallower component at $\sim +60$\,km\,s$^{-1}$ 
are visible in the lines of \ion{Ca}{2}\,H\&K. The \ion{Na}{1}\,D doublet 
exhibits a broad, saturated profile at almost zero velocity blended with a 
narrow component near 15\,km\,s$^{-1}$, plus another narrower component 
centered at $\sim 52$\,km\,s$^{-1}$. The reddest component displays a slightly 
higher velocity in \ion{Ca}{2}\,H\&K compared to the \ion{Na}{1}\,D doublet. 
The relatively simple line profiles of SN\,2021rhu do not indicate the presence 
of numerous resolved velocity components that typically span a velocity range 
of a few hundred km\,s$^{-1}$ as found in some strongly reddened Type Ia SNe 
(see, e.g., SN\,2006X, \citealp{Patat_etal_2007};
SN\,2014J, \citealp{Graham_etal_2015}; and 
SN\,1999cl, \citealp{Blondin_etal_2009}). SN\,2007le exhibited narrow, 
time-variable \ion{Na}{1}\,D absorption and a moderate amount of 
reddening \citep[$E(B-V)=0.27$\,mag;][]{Simon_etal_2009}. The same feature in 
SN\,2021rhu displayed a single component, while a complex absorption profile 
with $\sim 7$ distinct velocity components was found in SN\,2007le. Most of the 
\ion{Na}{1} and \ion{Ca}{2}\,H\&K profiles in SN\,2021rhu are confined between 
$\sim -20$ and $+60$\,km\,s$^{-1}$. They most likely arise in a single 
interstellar dust sheet intersecting the sight line. The lack of temporal 
evolution suggests constant ionization conditions of the dust cloud exposed to 
the variable SN radiation field between the observations of HIRES spectroscopy 
at days $+$39 and $+74$. However, the Keck HIRES observations starting from day 
$+$39 cannot rule out the possibility that the CSM may have been present at 
smaller distances to the SN and evaporated by the radiation at early phases.
Alternatively, the Na lines were absorbed by the foreground dust grains at 
large distances. In any case, we infer a rather clean circumstellar environment 
around day $+$80, and a large distance between SN\,2021rhu and the dust cloud 
where the interstellar lines form. \\

\section{Discussion} \label{sec:discussion}
The polarization of SN\,2021rhu shows a steep increase from the red to the 
blue wavelengths and peaked significantly blueward of the galactic average 
at $\approx$5500 \AA\ \citep{Whittet_etal_1992}, i.e., 
$p_{\rm max}=$1.78$\pm$0.02\% and $\lambda_{\rm max}=$3890$\pm$93 \AA. 
Such a behavior has been seen in some Type Ia SNe that show strong 
extinction ($E_{B-V}^{\rm host}\gtrsim$0.5 mag), but the peak polarization 
of SN\,2021rhu is lower compared to such events. For example, 
SNe\,1986G ($p_{\rm max}=$5.16$\pm$0.04\%, 
$\lambda_{\rm max}=$4300$\pm$10 \AA, \citealp{Hough_etal_1987}), 
2006X ($p_{\rm max} \textgreater$8\%, 
$\lambda_{\rm max}\lesssim$4000 \AA, \citealp{Patat_etal_2009}), 
2008fp ($p_{\rm max} \textgreater$2.2\%, 
$\lambda_{\rm max}\lesssim$4200 \AA, \citealp{Cox_Patat_2014}), and 
2014J ($p_{\rm max} \textgreater$6.6\%, 
$\lambda_{\rm max}\lesssim$4000 \AA, \citealp{Kawabata_etal_2014, 
Patat_etal_2015, Srivastav_etal_2016, Porter_etal_2016, Yang_etal_2018_pol}). 
However, the polarization wavelength dependence of SN\,2021rhu is still 
significantly deviated from the MilkyWay average, indicating an enhanced 
proportion of small grains along the Earth-SN\,2021rhu sightline than in 
the mean Galactic dust \citep{Whittet_etal_1992, Draine_2003a}.

After the ISP correction based on the low continuum polarization of Type Ia 
SNe around peak brightness, major spectral lines of SN\,2021rhu show moderate 
polarization.  This behavior is consistent with that generally found for Type 
Ia SNe. The low continuum polarization at day +$79$ also suggests the absence 
of a large amount of CSM within $\sim 2 \times 10^{17}$\,cm. This is because the 
photons scattered by a circumstellar dust cloud can be highly polarized at 
large scattering angles, causing a significant increase in the continuum 
polarization \citep{Wang_etal_1996, Yang_etal_2018_pol}. In order to produce 
a net polarized signal via CSM scattering, the CSM distribution on the plane 
of the sky needs to deviate from point symmetry. Such symmetry would lead to 
a complete cancellation of the electric vectors (i.e., zero net polarization). 
SN\,2021rhu exhibits a spherical explosion as indicated by the low continuum 
polarization that is consistent with zero as measured from as early as $\sim$seven 
days before the peak luminosity. Such a high degree of spherical symmetry is 
inconsistent with the WD-WD merger-induced explosion models, which predict 
modest ($\gtrsim$0.3) to strong ($\sim$1--2\%) continuum polarizations as seen 
along and perpendicular to the equatorial plane, 
respectively \citep{Bulla_etal_2016a}. The intermediate line polarization 
observed before and around the peak luminosity of SN\,2021rhu and the presence 
of the small-scale polarization structures across the spectral lines are 
compatible with the moderate chemical nonuniformity predicted by the 
delayed-detonation and sub-$M_{\rm Ch}$ Helium-shell detonation 
scenarios \citep{Bulla_etal_2016b}. 
Spectropolarimetry of Type Ia SNe beyond $\sim 30$\,days is still very rare. 
Our observations of SN\,2021rhu offer the first opportunity to follow the 
temporal evolution, and exploit the diagnostic power, of the polarization of 
spectral features at such late epochs.

\subsection{Multiple Polarization Components of \ion{Si}{2}\,$\lambda$6355~\label{sec:si}}
The additional higher-resolution spectropolarimetry obtained around the peak 
luminosity of SN\,2021rhu allows us to determine the geometric signatures left 
behind by the propagation of the burning front at smaller physical scales, 
on the order of a few hundred km\,s$^{-1}$. As shown in 
Figure~\ref{Fig_iqu_ep3}, the polarization modulation, which is unresolved in 
the conventional flux spectrum (although the latter has a spectral resolution 
that is more an order of magnitude higher), suggests that more than one major 
line-forming region is present. For comparison, in Figure~\ref{Fig_iqu_ep3} 
we also present the Stokes parameters obtained at the same epoch with the 
low-resolution 300V grism as shown in Figure~\ref{Fig_iqu_ep2}. Such a 
polarized line complex near the SN light-curve peak has also been reported for 
SN\,2018gv \citep{Yang_etal_2020}. The shape of the polarized line profile of 
SN\,2021rhu is not identical to that in SN\,2018gv, but both cases may arise 
from multiple opacity components of clumps and/or shells. 
Additionally, the small-scale modulations on the order of a few hundred 
km\,s$^{-1}$ can only be discerned in the spectropolarimetry spectra obtained 
with the higher-resolution grism 1200R, rather than the low-resolution grism 
300V. The comparison between the low- and high-resolution observations also 
demonstrates the feasibility of resolving small-scale structures in the SN 
ejecta with high-resolution spectropolarimetry. 

In addition to the sub-structures across the polarization spectrum of the 
\ion{Si}{2}$\lambda$6355 at the SN peak luminosity, polarization peaks are also 
seen across both \ion{Si}{2}$\lambda$6355 and \ion{Si}{2}$\lambda\lambda$5898, 
5979 absorption features at day $-$7. These narrow peaks are no longer 
distinguishable at day $+$0 (see Figures~\ref{Fig_iqu_ep1}--~\ref{Fig_iqu_ep2}, 
and Figure~\ref{Fig_pol}). The narrow polarization components are real and they 
are also seen under smaller sizes of spectral binning. Their FWHM widths are 
$\lesssim$1,000 km s$^{-1}$ as measured at day $-$7, reflecting the presence 
of Si-rich clumps with a similar scale in velocity space that intersect the 
photosphere. At day $+$0, the photosphere recedes into the deeper layers of the 
ejecta. The disappearance of these narrow polarization peaks and the evolution 
of the small-scale modulations seen from Figures~\ref{Fig_iqu_ep2}, 
~\ref{Fig_pol}, and ~\ref{Fig_iqu_ep3} can be understood as the retreating 
photosphere has passed the clumps of Si-line opacities that produced the narrow 
polarization peaks at day $-$7. Future spectropolarimetric observations with a 
higher cadence will provide a greater spatial resolution in depth, therefore 
enabling a more comprehensive tomography of the SN ejecta and trace the 
geometric properties of any small-scale structures.

\subsection{The Dominant Axes of the Continuum and the \ion{Ca}{2}\,NIR3 Line  Polarization~\label{sec:qu}}
Presenting the spectropolarimetry on the Stokes $Q$--$U$ plane provides an 
intuitive layout of the axisymmetry of the continuum and across different 
spectral features \citep{Wang_etal_2001}. For example, an axially symmetric 
structure will imprint a straight line in the Stokes $Q$--$U$ plane, the 
so-called ``dominant axis'' \citep{Wang_etal_2003_01el, Maund_etal_2010_05hk}: 
\begin{equation}
U = \alpha + \beta Q. 
\label{Eqn_daxis}
\end{equation}
After ISP correction, departures from spherical and axial symmetries in the 
ejecta are indicated in the $Q$--$U$ diagram by their distances from the origin 
and deviations from the dominant axis, respectively. 

\begin{figure}[h]
\epsscale{1.18} 
\plotone{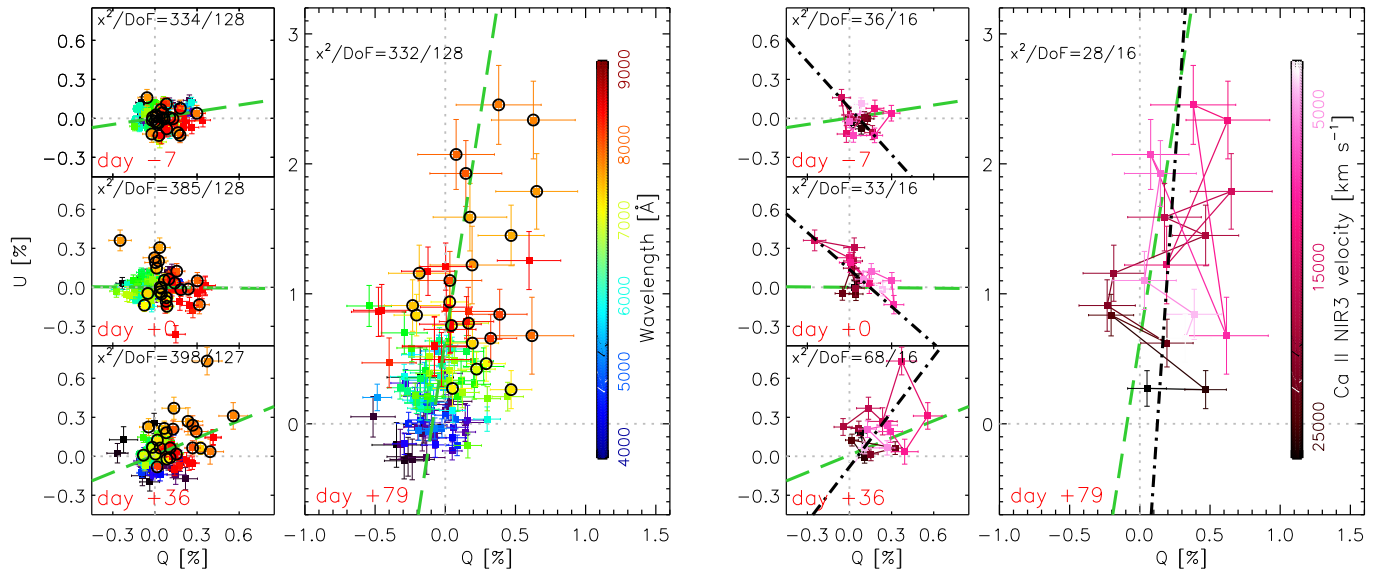}
\caption{ISP-corrected observations of SN\,2021rhu in the Stokes $Q$--$U$ plane. 
The data have been rebinned to 40\,\AA. The four subpanels on the left (three small square panels for epochs 1--3, plus a larger rectangular panel for epoch 
4) present the data in the range 3850--9100\,\AA, the wavelengths of which are indicated 
by the left color bar. The ranges of $Q$ and $U$ in the square and rectangular 
panels are different, but the scales for $Q$ and $U$ are the same in all 
subpanels. The long green long-dashed lines trace the dominant axis of the 
continuum at different epochs. Note that the dominant axis fitting at day $+$79 
does not represent the orientation of the optical ejecta (see 
Sec.~\ref{sec:qu}). Data points marked with open black circles indicate the 
spectral region covering the \ion{Ca}{2}\,NIR3 feature. The corresponding four 
subpanels on the right show the \ion{Ca}{2} line over the velocity range 
27,000--2,000\,km\,s$^{-1}$. The right color bar indicates the velocity range of the \ion{Ca}{2}\,NIR3 complex. The light-green long-dashed lines in both 
left and right subpanels represent the dominant axis fitted to the presented 
optical wavelength range, while the gray dot-dashed lines in the right 
subpanels trace the dominant axis over the \ion{Ca}{2}\,NIR3 feature. 
\label{Fig_qu}
}
\end{figure}

In Figure~\ref{Fig_qu} we present the ISP-corrected Stokes parameters in the 
Stokes $Q$--$U$ plane between days $-$7 and $+$79. The dominant axes of 
SN\,2021rhu were determined by performing an error-weighted linear 
least-squares fitting to the continuum polarization 
($3850 \leq\lambda\leq 9100$\,\AA) and to the \ion{Ca}{2}\,NIR3 feature 
($7800 \lesssim\lambda\lesssim 8510$\,\AA, corresponding to a velocity range 
from 27,000 to 2000\,km\,s$^{-1}$), respectively. The fitted dominant axes 
for SN\,2021rhu at different epochs are plotted in Figure~\ref{Fig_qu}, and 
the derived parameters are given in Table~\ref{Table_pol}. Based on the 
$Q$--$U$ diagrams in the left panels of Figure~\ref{Fig_qu}, the optical ejecta 
of SN\,2021rhu observed in the first three epochs (days $-$7 to $+$36) are 
consistent with a dominant axis that rotates with time, while deviations from 
a single axial symmetry are also notable as indicated by the scatter about 
the dominant axis reflected by the large values of $\chi^2$ labeled in each 
subpanel. At day $+79$, the dominant axis fitted to the optical spectrum is 
mostly defined by the significantly polarized \ion{Ca}{2}\,NIR3 line. After 
excluding the wavelength range covering the \ion{Ca}{2}\,NIR3 feature (i.e., 
the black-circled data points in the left rectangular panel of 
Fig.~\ref{Fig_qu}), the remaining data points are scattered around (0,0) 
and do not exhibit any prominent dominant axis. Therefore, we conclude that 
the optical ejecta do not follow any conspicuous axial symmetry at day $+79$. 

The $Q$--$U$ diagram across the \ion{Ca}{2}\,NIR3 feature exhibits complicated configurations. Loop-like structures can be seen before day $+$36. Such 
patterns across the \ion{Ca}{2}\,NIR3 line are observed in other typical Type 
Ia SNe and represent variations in the amplitude and orientation of the 
polarization as a function of velocity or depth, indicating the deviation 
from axial symmetry \citep{Wang_wheeler_2008}. At the last epoch at day $+$79 
(the large left and right subpanels in Fig.~\ref{Fig_qu}), the scatter about 
the dominant axis of the Ca line is smaller, but the complex \ion{Ca}{2}\,NIR3 
polarization profile cannot be described by a single loop. Within the 
framework of a model in which the polarization is produced by line opacity 
that unevenly blocks the underlying photosphere, the \ion{Ca}{2}\,NIR3 
polarization may indicate the presence of multiple Ca-rich components in the 
inner layers of the ejecta. Although the photosphere of Type Ia SNe may still 
persist after day $\sim 100$ as evidenced by the presence of permitted 
lines \citep{Black_etal_2016}, the above interpretation of the presence of 
multiple Ca-rich components obscuring the SN photosphere will be invalid if, 
at late phases, the ejecta have become sufficiently optically thin that 
electron scattering may not be able to produce considerable polarization 
signals. In \S \ref{sec:gsa}, we propose an alternative mechanism of atomic 
alignment that generates polarized signals not by patchy photospheric 
electron scattering but through the alignment of atomic angular momentum in an 
anisotropic radiation field. 


\subsection{The Late-Time Increase in the Polarization of the \ion{Ca}{2} 
Lines~\label{sec:capol}}
The apparent increasing \ion{Ca}{2}\,NIR3 polarization of SN\,2021rhu at days 
$+$36 and $+$79 has previously only been reported for SN\,2006X, which showed 
an increase from $\sim 0.6$\% around maximum luminosity to $\sim 1.2$\% 
at day $+$39 \citep{Patat_etal_2009}. SN\,2001el, which is the only other Type 
Ia SN with \ion{Ca}{2}\,NIR3 polarization measured at a considerable late 
phase, did not show such polarization signals at day 
$+$41 \citep{Wang_etal_2003_01el}. Although the uncertainty becomes larger 
near the blue end of the spectral coverage, a similar evolution of \ion{Ca}{2} 
polarization is also seen from the \ion{Ca}{2}\,H\&K lines of SN\,2021rhu. The 
nondetection of the \ion{Ca}{2}\,NIR3 polarization in SN\,2001el at day 
$+$41 \citep{Wang_etal_2003_01el} may be due to an orientation effect or an 
intrinsic diversity among Type Ia SNe. Considering the sparse sample of 
polarimetry, a meaningful conclusion is not currently feasible. 

The evolution of the \ion{Ca}{2}\,NIR3 from $\sim 0.3$\% at around the peak to 
$\sim 2.5$\% at day +79 is thus unprecedented. For example, the polarization 
peak in the \ion{Ca}{2}\,NIR3 profile of Type Ia SNe around maximum light is 
generally $\lesssim 1$\% \citep{Wang_wheeler_2008}. By far the only exception, 
SN\,2004dt, displayed a peak \ion{Ca}{2}\,NIR3 polarization of 
$\sim 0.7$--2\% \citep{Wang_etal_2006_04dt}. This high line polarization was 
observed around the peak luminosity, and it is incompatible with the 
$\lesssim 0.8$\% maximum line polarization predicted by hydrodynamic models 
for two-dimensional (2D) double detonation and three-dimensional (3D) delayed 
detonation \citep{Bulla_etal_2016b}. 

This ``repolarization" behavior seen in SN\,2006X has been explained in the 
context of a deflagration/detonation model by a partial blocking of the 
photosphere when it had receded to within the inner edge of the Ca layer at 
8000--9000\,km\,s$^{-1}$ \citep{Patat_etal_2009}. Detailed modeling of the 
degree of asymmetry between the bottom of the Ca layer and the outer parts of 
the Fe-rich ejecta at later times is necessary to determine whether the 
\ion{Ca}{2}\,NIR3 polarization of SN\,2021rhu at day $+$79 can be accommodated 
in the context of an $M_{\rm Ch}$ deflagration/detonation picture. This may be 
a challenge. As the ejecta expand, the optical depth decreases approximately 
$\propto r(t)^{-2}$ or $\propto t^{-2}$, where $r(t)$ represents the radius 
reached by freely expanding ejecta at time $t$ after the SN explosion. 
After a few months, the size of the $\tau=2/3$ photosphere has 
substantially decreased. The light emitted by the SN is dominated by numerous 
overlapping line transitions of Fe-group elements. The abundant overlying 
narrow lines form a `quasi-continuum' spectrum, superimposed by several strong 
Fe and Co emissions. The late-time emission-dominated spectra trace the 
distribution of the Fe-group burning products near the central energy source. 
These emissions are essentially unpolarized, resulting in an absence of 
polarized photons to be blocked in the first place. 
Therefore, the decreasing scattering cross-section over time makes the 
photosphere-obscuration mechanism less likely to account for the significant 
\ion{Ca}{2} polarization at late times. Owing to the relatively large 
systematic uncertainties in the very blue end of the optical continuum 
(\ref{sec:fors2}), we refrain from an interpretation of the polarization 
behavior measured across the \ion{Ca}{2}\,H\&K lines.

\subsection{Considerations Based on Atomic Alignment in a Weak Magnetic Field~\label{sec:gsa}}
We suggest an alternate possibility that might produce the high \ion{Ca}{2} 
polarization as found in SN\,2021rhu at day $+79$. This process involves 
photo-excitation in an anisotropic radiation field when the 
electron-scattering opacity in the SN ejecta has decreased substantially. 
The lower levels of \ion{Ca}{2}\,NIR3 are metastable states that can be 
geometrically aligned through photo-excitation by an anisotropic radiation 
field. In an interstellar medium that has a weak magnetic field, a subsequent 
realignment of the angular momentum of the atoms in their ground state may 
happen through magnetic precession (see, e.g., \citealp{Happer_1972,
Landolfi_etal_1986}). Such a magnetic realignment will take place if the 
Larmor precession rate, $\nu_{\rm L}$, is greater than the photo-excitation 
rate, $\tau_{R}^{-1}$, from the ground state of the atoms: 
$\nu_{\rm L} \textgreater \tau_{R}^{-1}$ \citep{Yan_Lazarian_2006}. The atoms' 
angular momentum will then be realigned with respect to the magnetic field. 
In the case of $\nu_{\rm L} \approx \tau_{R}^{-1}$, this ``Hanle effect" will 
become effective in a relatively stronger magnetic field \citep{Hanle_1924, 
Ignace_etal_1997, Yan_Lazarian_2008}. 

The incident flux must be anisotropic in order to differentially excite the 
atoms in different magnetic sublevels (see, e.g., \citealp{Happer_1972,  
Landolfi_etal_1986}; and the review by \citealp{Yan_Lazarian_2012}). In the 
presence of anisotropic flux, the angular momentum of the atoms will also be 
distributed anisotropically. The result is the induction of unequal 
populations over the magnetic sublevels that correspond to different magnetic 
quantum numbers, and hence the production of polarized radiation. This picture 
is compatible with the configuration of Type Ia SNe in which \ion{Ca}{2}-rich 
matter is illuminated by a central emission source. In this case, the 
radiation field is primarily radial and hence naturally anisotropic, and the 
photon pumping is intrinsically anisotropic. Under the framework of atomic 
alignment, the spatial distribution of \ion{Ca}{2} cannot be inferred from its 
polarization. The photons are polarized through the interaction between 
optical pumping by an anisotropic radiation field and the ambient magnetic 
field, so the polarization mechanism will be in effect regardless of the 
spatial distribution of \ion{Ca}{2}. This is totally different from the 
photosphere-blocking mechanism.

Particularly in the later stages of the expansion of a Type Ia SN, the ejecta 
become thin enough that the impact of collisions diminishes. In the case of 
\ion{Ca}{2}, the photo-excitation from the ground state $^2S_{1/2}$ is 
dominated by the two E1 transitions of the \ion{Ca}{2}\,H$\&$K lines and 
followed by the cascade to the two metastable states $^2D_{3/2,5/2}$. As 
proposed by \cite{Yan_Lazarian_2006}, the metastable states of 
\ion{Ca}{2}\,NIR3 can be aligned in an anisotropic radiation field and 
magnetically realigned in the same fashion as their ground states. This 
happens to some atomic species because their metastable states are also 
long-lived, which makes them act similarly to the ground states and become 
sensitive to a weak magnetic field. 



Assuming a black-body radiation field of 5000\,K, the photo-excitation rate 
from the ground state of \ion{Ca}{2}\,NIR3 gives a photo-excitation rate of 
$\sim2\times 10^4$\,s$^{-1}$ \citep{Yan_Lazarian_2006}. This corresponds to the 
inverse of the lifetime of the metastable states, which in turn yields a low 
magnetic sensitivity of $\sim1$\,mG for the 
\ion{Ca}{2}\,NIR3 \citep{Carlin_etal_2013}. In other words, the polarization 
of the \ion{Ca}{2}\,NIR3 line potentially traces the component of the local 
magnetic field in the plane of the sky when the magnetic field is stronger 
than $\sim 1$\,mG. Based on the optical pumping from the ground state, 
particularly, we estimated the maximum degree of polarization for 
\ion{Ca}{2}\,NIR3 reached in the ideal alignment case, with a beam of 
radiation and a local magnetic field aligned with it. Without counting the 
effect of collisions between atoms, the maximum degree of polarization for 
8500.36\,\AA\ ($2D_{3/2} \textgreater 2P_{3/2}$), 8544.44\,\AA\ ($2D_{5/2} \textgreater 2P_{3/2}$), and 8664.52\,\AA\ ($2D_{3/2} \textgreater 2P_{1/2}$)
can reach 10.8\%, 8.9\%, and 12.5\%, respectively. 
The highest polarization is higher compared to that in \ion{Ti}{2} lines, 
$\sim 7$\% for the two 
aligned metastable ($2D_{3/2, 5/2}$) to the upper ($2P^{O}_{1/2, 3/2}$) states, 
which exhibits similar but more complicated transitional structures compared to 
\ion{Ca}{2}\,NIR3 (see, e.g., Table~3 of \citealp{Yan_Lazarian_2012}). A more 
detailed discussion of the \ion{Ca}{2}\,NIR3 polarization computed based on the 
ground-state-alignment theory will be presented in future work (Yan et al., in 
prep.). 

Since all three components of the \ion{Ca}{2}\,NIR3 triplet are magnetically 
alignable and can be individually polarized, the large width of the polarized 
feature on day $+$79 (see Figs.~\ref{Fig_iqu_ep5}, \ref{Fig_pol}, and 
\ref{Fig_qu}) is in agreement with the proposed atomic alignment mechanism. 
The polarization profile and peak level at this late time differ 
substantially from those around peak luminosity (see 
Figs.~\ref{Fig_iqu_ep1}--\ref{Fig_iqu_ep2}), when the line polarization 
was mainly due to an uneven obscuration of the photosphere by the line opacity. 
If the late-time polarization is indeed caused by atomic alignment in a weak 
magnetic field, the shape of the polarization profile may depend on (i) the 
anisotropy of the pumping radiation  field, (ii) the ambient magnetic field, 
and (iii) the velocity field that affects the Ca opacity. The modeling of the 
\ion{Ca}{2}\,NIR3 polarization profile is beyond scope of this paper. 

For atomic species to be aligned, the angular momentum of their ground state 
should be nonzero. For example, \ion{Ca}{2}\,H\&K, which is induced via the 
transition from the ground state to the upper state (i.e., 
$2S_{1/2}\rightarrow2P^{O}_{1/2, 3/2}$), is not alignable by this magnetic 
mechanism since the total angular momentum of the ground state is $J=0$, which
does not allow uneven occupation of ground-state angular momenta. Therefore, 
the \ion{Ca}{2}\,H\&K absorption would not be polarized by the same effect. 
However, although the uncertainty in the Stokes parameters becomes larger near 
the blue end of the spectral coverage, a similar evolution of increasing 
late-time \ion{Ca}{2} polarization is also seen in the \ion{Ca}{2}\,H\&K lines. 
In Figures~\ref{Fig_iqu_ep4} and \ref{Fig_iqu_ep5}, we present the observed 
polarization position angles across both the \ion{Ca}{2}\,H\&K and 
\ion{Ca}{2}\,NIR3 for SN\,2021rhu after correcting for the ISP. At day $+$79, 
the two position angles exhibit significant differences, suggesting that the 
polarization mechanisms of the two \ion{Ca}{2} lines are intrinsically 
different. In the case of uneven photosphere obscuration, the two features 
would be expected to exhibit similar polarization position angles, since both 
lines are likely to originate from the same \ion{Ca}{2} opacity distribution. 


\section{Summary}
Compared to the time around peak luminosity, we found a strong growth of the 
\ion{Ca}{2} polarization observed in the Type Ia SN\,2021rhu on days $+$36 and 
$+$79. The continuum polarization remained low at day $+$79 in spite of the 
drastic increase in polarization of \ion{Ca}{2}\,NIR3; this is consistent with 
the low level measured for the continuum polarization at early phases when the 
photosphere has not yet receded into deep layers. We consider the possibility 
of line polarization owing to partial blocking of the underlying photosphere by 
Ca-rich material, and an alternative explanation that \ion{Ca}{2}\,NIR3 might 
be polarized through the alignment of the atoms with respect to an ambient 
weak magnetic field in an anisotropic radiation field. Detailed modeling of 
the late-time polarization of \ion{Ca}{2} features by magnetic alignment 
should consider (1) the geometric distribution of the \ion{Ca}{2} opacities, 
(2) the shape and degree of anisotropy of the induced radiation field, and 
(3) the strength and geometry of the magnetic field. Note that the radiation 
field above the photosphere is intrinsically anisotropic, and the majority of 
the photon pumping will happen along the radial directions. Such an anisotropic 
incident flux can induce an unequal population distribution over the magnetic 
sublevels. Without the flux anisotropy, the alignment of atomic angular 
momentum will not happen, and no polarization will be produced. 

Assuming that SN\,2021rhu is not a singular case, we briefly outline the open 
questions raised by the polarimetry of SN\,2021rhu at late phases. \\
(i) Is the large \ion{Ca}{2}\,NIR3 polarization observed on day $+$79 best explained by photospheric blocking or optical pumping effects or their combination, or is another process needed? \\
(ii) Do different explosion mechanisms predict specific late-time continuum and line polarization properties that can be used for diagnostic purposes? \\
(iii) Can the magnitude and geometry of the magnetic field be determined from the late-time \ion{Ca}{2}\,NIR3 polarization profile? \\
(iv) Can the multicomponent polarization profiles of \ion{Si}{2}\,$\lambda$6355 and \ion{Ca}{2}\,NIR3 and their evolution lay the foundation for an overarching consistent model? 

During the rapid expansion of the ejecta, any pre-existing magnetic field will 
be frozen in the ejecta and weaken as they expand. Evidence for high initial 
magnetic fields (in excess of 10$^{6}$\,G at the surface of the progenitor 
WD) has been reported from infrared line profiles and light curves of Type Ia SNe 
(see, e.g., \citealp{Penney_etal_2014, Hristov_etal_2021}). The exact nature of 
the initial magnetic field of the WD and its evolution to late SN phases may be 
of great importance for the understanding of Type Ia SNe. Owing to the rapid 
evolution of the SN ejecta, the radiation field may become sufficiently diluted 
and anisotropic, matching the conditions necessary for an effective atomic 
alignment for a wide range of atomic species and transitions. A rise of 
polarized atomic lines is expected at this epoch. Perhaps this mechanism is 
already at work in the observed polarization of \ion{Si}{2} and \ion{Ca}{2} 
lines even around maximum brightness. 

This is a new area worth pursuing both theoretically and observationally in 
future studies. With a comprehensive understanding of the mechanism(s), the 
\ion{Ca}{2} polarization may provide a measure in the ejecta of the magnetic 
field inherited from the progenitor. This would be a valuable new diagnostic 
of Type Ia SN explosion models. Future spectropolarimetry of Type Ia SNe 
extending to late phases is also essential to search for any systematics, to 
explain the polarization mechanism, to discriminate between orientation and 
intrinsic diversity, and to further understand the radiation distribution 
within the core. 

\setlength{\tabcolsep}{3.5pt}
\begin{table}[!h]
\begin{center}
\caption{VLT spectropolarimetic observations of SN\,2021rhu. \label{Table_pol}}
\begin{scriptsize}
\begin{tabular}{c|ccc|cc|cc|ccc}
\hline
\hline
\#  &    Date (UT) /     &  Exp. Time$^b$  & Grism /       & $q^{\rm Cont}$ & $u^{\rm Cont}$ &  $\alpha$ (\%)         &  $\theta_{d}$    &  $\alpha^{\rm Ca\,II\,NIR3}$ (\%)   &  $\theta_{d}^{\rm Ca\,II\,NIR3}$   &  $p_{\rm max}^{\rm Ca\,II\,NIR3}$  \\

       &   Phase$^a$ (day)  &  (s)        &  Res.\ Power  & [\%] &  [\%]        &    $\beta$                &  (deg)           &  $\beta^{\rm Ca\,II\,NIR3}$            &  (deg)          &  (\%)                              \\

\hline
1      &   2021-07-08 /     &  $2\times4\times150$  &     300V/440 & 0.04$\pm$0.12 & 0.01$\pm$0.11 &  0.00322$\pm$0.00391   &  184.7$_{-1.1}^{+1.1}$   &  0.0795$\pm$0.0765               &  155.0$_{-7.1}^{+16.4}$  &  0.30$\pm$0.06  \\
       &   $-$7               &            &             & & &    0.166$\pm$0.040    &       &  $-$1.194$\pm$0.885          &           &    \\
\hline

2      &   2021-07-15 /     &  $2\times4\times80$  &     300V/440  & 0.00$\pm$0.12 & 0.00$\pm$0.11 &  9.60$\times10^{-5}\pm$0.00354   &  179.7$_{-1.3}^{+1.3}$   &  0.138$\pm$0.029               &  158.2$_{-3.2}^{+4.0}$  &  0.44$\pm$0.08  \\
       &   $+$0             &              &           & & &   $-$0.0107$\pm$0.0439    &       &  $-$0.951$\pm$0.238        &           &    \\
\hline

2$^{c}$   &   2021-07-15      &  $2\times4\times160$  &    1200R/2140  & & &  N.A.  &  N.A.   &  N.A.  &  N.A.  &  N.A.  \\
\hline

3      &   2021-08-20 /     &  $2\times4\times450$   &     300V/440 & -0.03$\pm$0.12 & 0.00$\pm$0.12 &  0.00894$\pm$0.00516   &  191.9$_{-1.6}^{+1.5}$   &  -0.0817$\pm$0.0891               &  207.3$_{-5.1}^{+3.4}$  &  0.82$\pm$0.11  \\
       &   $+$36             &             &            & & &   0.441$\pm$0.063    &       &  1.412$\pm$0.430        &           &    \\
\hline

4      &   2021-10-02 /     &  $2\times4\times300$   &     300V/440 & -0.11$\pm$0.38 & 0.14$\pm$0.41 &  N.A.           &  N.A.            &  $-$1.664$\pm$7.104               &  222.2$_{-25.6}^{+1.3}$  &  2.48$\pm$0.31  \\
       &   $+$79             &               &          & & &   N.A.    &       &  1.412$\pm$0.430        &           &    \\
\hline
\end{tabular}\\
{$^a$Relative to the estimated peak on UT 2021-07-15/MJD 59410.} \\
{$^b$Each set of observation consists of 2 [loops]$\times$4 [half-wave plate angles]$\times$[time of integration].} \\
{$^c$A higher spectral resolution and a narrower wavelength range of 5700--7100\,\AA\ are offered by the grism 1200R.}
\\
\end{scriptsize}
\end{center}
\end{table}

\begin{acknowledgments}
We thank the anonymous referee for the careful scrutiny which resulted in 
quite a few helpful suggestions that improved the quality of the manuscript. 
We are grateful to the European Organisation for Astronomical Research 
in the Southern Hemisphere (ESO) for the generous allocation of observing 
time. The polarimetry studies in this work are based on observations made 
with the VLT at ESO's La Silla Paranal Observatory under program ID 
105.20AU. We especially thank the staff at Paranal for their proficient 
and highly motivated support of this project in service mode. 
The high-resolution spectra presented herein were obtained at the 
W.\,M.\,Keck Observatory, which is operated as a scientific partnership 
among the California Institute of Technology, the University of California, 
and the National Aeronautics and Space Administration (NASA). The Observatory was 
made possible by the generous financial support of the W.\,M.\,Keck 
Foundation. 
The authors wish to recognize and acknowledge the very significant 
cultural role and reverence that the summit of Maunakea has always had 
within the indigenous Hawaiian community. We are most fortunate to have 
the opportunity to conduct observations from this mountain.

PyRAF, PyFITS, STSCI$\_$PYTHON are products of the Space Telescope Science 
Institute, which is operated by AURA for NASA. STScI is operated by the 
Association of Universities for Research in Astronomy, Inc., under NASA 
contract NAS5-26555. This research has made use of NASA's Astrophysics 
Data System Bibliographic Services, the SIMBAD database, operated at CDS, 
Strasbourg, France, and the NASA/IPAC Extragalactic Database (NED) which 
is operated by the Jet Propulsion Laboratory, California Institute of 
Technology, under contract with NASA. 

The research of Y.Y.\ is supported through a 
Bengier-Winslow-Robertson Fellowship. 
M.B.\ acknowledges support from the Swedish Research Council (Reg. No. 2020-03330). 
A.V.F.'s group at U.C.\ Berkeley acknowledges generous support from the Miller Institute for Basic Research in Science (where A.V.F. was a Miller Senior Fellow), Sunil Nagaraj, Landon Noll, Gary and Cynthia Bengier, Clark and Sharon Winslow, Sanford Robertson, and many additional donors. 
L.G.\ acknowledges financial support from the Spanish Ministerio de Ciencia e Innovaci\'on (MCIN), the Agencia Estatal de Investigaci\'on (AEI) 10.13039/501100011033, and the European Social Fund (ESF) ``Investing in your future" under the 2019 Ram\'on y Cajal program RYC2019-027683-I and the PID2020-115253GA-I00 HOSTFLOWS project, from Centro Superior de Investigaciones Cient\'ificas (CSIC) under the PIE project 20215AT016, and the program Unidad de Excelencia Mar\'ia de Maeztu CEX2020-001058-M. 
P.H.\ acknowledges the support from the NSF project ``Signatures of Type Ia Supernovae, New Physics, and Cosmology,'' grant AST-1715133. The supernova research by L.W.\ is supported by NSF award AST-1817099. M.R.\ is supported by the National Science Foundation Graduate Research Fellowship Program under grantr DGE-1752134. 
J.C.W.\ and J.V.\ are supported by NSF grant AST-1813825. 
The research of J.M.\ is supported through a Royal Society University 
Research Fellowship. 
  
\end{acknowledgments}

\vspace{5mm}
\facilities{VLT(FORS2), Keck:I (HIRES)}


\software{IRAF \citep{Tody_1986, Tody_1993}
          }

\bibliographystyle{aasjournal}

%





\end{document}